# Two-dimensional antiferromagnets with non-relativistic spin splitting switchable by electric polarization

Himanshu Mavani, Kai Huang, Kartik Samanta, and Evgeny Y. Tsymbal*

*Department of Physics and Astronomy & Nebraska Center for Materials and Nanoscience,*
*University of Nebraska, Lincoln, Nebraska 68588-0299, USA*

**ABSTRACT**. Spin-split antiferromagnets have significance for antiferromagnetic (AFM) spintronics due to their momentum dependent spin polarization which can be exploited for the control and detection of the AFM order parameter. Here, we explore the polar-layer stacking of AFM-ordered bilayers driving the emergence of reversable electric polarization and non-relativistic spin splitting (NRSS) of their band structure. Based on the spin-space group approach, we identify several representative two-dimensional AFM materials which exhibit different types of NRSS when stacked into a polar bilayer. We demonstrate that NRSS can have both altermagnetic and non-altermagnetic origins and elucidate symmetry requirements for NRSS to be switchable by electric polarization. We argue that the electric polarization switching of NRSS in polar AFM bilayers may be more practical for device applications than the current-induced Néel vector switching.

*Keywords*: altermagnetism, sliding ferroelectricity, 2D van der Waals material, polar stacking

* E-mail: tsymbal@unl.edu

## I. INTRODUCTION

Antiferromagnetic (AFM) spintronics has recently emerged as a subfield of spintronics, where an AFM order parameter serves as a state variable [1]. Due to being robust against magnetic perturbations, producing no stray fields, and exhibiting ultrafast spin dynamics, antiferromagnets can serve as functional materials for spintronic applications [2]. Especially attractive are AFM materials exhibiting nonrelativistic spin splitting (NRSS) of their electronic band structure. In contrast to conventional collinear antiferromagnets with spin-degenerate bands enforced by $PT$ or $U\tau$ symmetries (where $P$ is space inversion, $T$ is time reversal, $U$ is spin flip, and $\tau$ is lattice translation), NRSS antiferromagnets break these symmetries producing momentum-dependent spin polarization, even in the absence of spin-orbit coupling (SOC) [3-5]. Among collinear NRSS antiferromagnets are those in which two opposite magnetic sublattices are connected by proper or improper spatial rotations, now known as altermagnets [6,7], and those in which the two opposite magnetic sublattices are not connected by any symmetry [8]. Due to their momentum-dependent spin polarization, NRSS antiferromagnets allow phenomena traditionally associated with ferromagnets, such as spin-polarized currents [9,10], exchange torques [11], anomalous Hall effects [12-14], spin filtering [15], and magnetoresistive effects [16-19], which make them appealing for spintronics. Based on spin-space group theory and *ab-initio* studies, a few altermagnetic materials have been predicted in three-dimensional (3D) [20] and two-dimensional (2D) systems [21-23], and some of them have been experimentally confirmed [24-29].

For the use of NRSS antiferromagnets in spintronics, switching of the AFM order parameter, known as the Néel vector, is considered as the necessary requirement. While it is possible to reverse the Néel vector using spin-transfer or spin-orbit torques [30], this approach is, in general, energy inefficient due to a large current density required to produce these torques [31]. Recently, the electric-field-induced NRSS has been proposed, where NRSS emerges due to applied electric field breaking inversion symmetry [32,33]. This approach is interesting because it allows the control of NRSS without the reversal of the AFM Néel vector and hence it is expected to be less power demanding. However, it suffers from a relatively small NRSS and the lack of non-volatile performance, which is desirable from a practical perspective, as well as appealing from the fundamental point of view.

These deficiencies can be overcome using polar-layer stacking of 2D antiferromagnets. Stimulated by the recent advances in van der Waals (vdW) assembly techniques [34], new ways to generate altermagnetism in 2D vdW systems have been proposed theoretically [35-39]. Especially interesting is polar-layer stacking, where two nonpolar vdW layers are arranged into non-centrosymmetric stacking configuration generating out-of-plane electric polarization [40]. Due to the polarization being reversable via layer sliding, this phenomenon was dubbed "sliding ferroelectricity" [41]. The experimental realization of sliding ferroelectricity has been demonstrated using BN, a vdW material which is non-polar in the bulk form [42-44], thus showing the practical relevance of this technique. Stimulated by these findings, several proposals have been made to realize NRSS in AFM vdW bilayer systems [45-47].

In this letter, we explore the effect of polar-layer stacking on NRSS in bilayer antiferromagnets and demonstrate that the spin splitting can have both altermagnetic and non-altermagnetic origin depending on the type of stacking. Based on spin-space group approach, we identify several representative 2D AFM vdW materials which exhibit different types of NRSS and elucidate symmetry requirements for NRSS to be switchable by electric polarization. We argue that the electric polarization switching of NRSS in polar AFM bilayers may be more practical for device applications than the Néel vector switching.

## II. METHODS

First-principles calculations are performed using the Vienna *ab initio* Simulation Package (VASP) [48]. The projector augmented wave (PAW) method is employed to describe interactions between ions and valence electrons, and the Perdew–Burke–Ernzerhof (PBE) functional is used to treat the exchange-correlation energy [49, 50]. A plane-wave cutoff energy of 500 eV and a 16×16×1 k-point mesh in the irreducible Brillouin zone are used in the calculations. Atomic positions are relaxed until the force on each atom is less than $2\times10^{-3}$ eV/Å, and self-consistency of electronic structure calculations is achieved with tolerance of $10^{-7}$ eV. An on-site Coulomb interaction is treated within a rotationally invariant DFT+U approach [51] using a Hubbard parameter $U = 4$ eV applied to the Mn $d$-orbitals of MnPSe$_3$ and MnPS$_3$ [52] and $U = 2$ eV applied to the Cr $d$-orbitals of 1T-CrTe$_2$. In the calculations of 2D systems, periodically repeated slabs are separated by a vacuum layer of more than 20 Å to avoid interactions between periodic images. Van der Waals interactions are included in the calculations using the semiempirical DFT-D3 method [53]. The out-of-plane electric polarization of polar-stacked bilayers is calculated using the dipole correction method [54]. Calculated lattice constants and interlayer distances of bilayers MnPSe$_3$, MnPS$_3$, and 1T-CrTe$_2$ are presented in Table S1 of Supplemental Material [55]. Several computational tools, including Bilbao Crystallographic Server [56], FINDSYM [57], Pymatgen [58], and PyProcar [59], are used for crystallographic analysis and electronic structure visualization.

## III. RESULTS AND DISCUSSION
### A. Symmetry analysis

To explore NRSS in bilayer antiferromagnets, we employ the spin-space group approach where the spin and space degrees of freedom are decoupled [6]. The spin-space group symmetry operation is represented by $[R_s||R_l]$, where the transformation on left (right) of the double bar acts only on the spin (space). Any collinear magnet holds spin-only symmetry $[\bar{C}_2||T]$ (where $\bar{C}_2$ is twofold spin rotation perpendicular to the collinear spin axis, $C_2$, followed by spin inversion), which makes electronic bands at momentum $\mathbf{k}$ and $-\mathbf{k}$ to have the same energy $\varepsilon$, as it transforms $\varepsilon(s,\mathbf{k})$ to $\varepsilon(s,-\mathbf{k})$. In 3D conventional collinear magnets, the spin degeneracy of electronic bands is protected by $[C_2||\bar{E}]$ and/or $[C_2||\tau]$ symmetries (where $\bar{E}$ is inversion). Operation $[C_2||\bar{E}][\bar{C}_2||T]$ is equivalent to $PT$, and $[C_2||\tau]$ is equivalent to $U\tau$, both transforming $\varepsilon(s,\mathbf{k})$ to $\varepsilon(-s,\mathbf{k})$ and thus enforcing Kramers' spin degeneracy. In 2D collinear magnets, the spin degeneracy is also protected by $[C_2||M_z]$ or $[C_2||C_{2z}]$ symmetries (e.g., $[C_2||C_{2z}][\bar{C}_2||T]$) transforms $\varepsilon(s,k_x,k_y)$ to $\varepsilon(-s,k_x,k_y)$). Thus, to generate NRSS in a 2D antiferromagnet, two magnetic sublattices cannot be connected by translation $\tau$, inversion $\bar{E}$, mirror reflection $M_z$, or twofold rotation $C_{2z}$. In a 2D altermagnet, however, they can be swapped by other crystal rotational symmetries (proper or improper), reflecting the alternating character of NRSS in the reciprocal space. NRSS can also have a non-altermagnetic character, if there are no crystal rotational symmetries connecting the two magnetic sublattices.

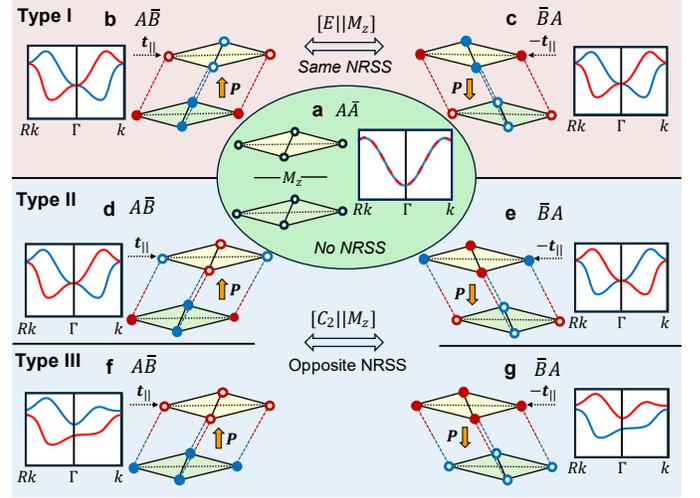

**FIG. 1**: Construction of polar-stacked AFM bilayer. (a) Stacking $A\bar{A}$ (left), where top monolayer $\bar{A}$ represents mirror reflection $M_z$ of bottom monolayer $A$, and corresponding spin-degenerate band structure (right). $k$ and $Rk$ denote wave vectors connected by rotational symmetry $R$. Red and blue lines denote spin-up and spin-down bands. (b), (d), (f) The up polarization and (c), (e), (g) down polarization ground states of polar-stacking bilayer. Red and blue circles denote two magnetic sublattices. Open and solid circles denote two different sizes of magnetic moments (demonstrating magnetoelectric effect of electric polarization). Type I, II, and III shows three possibilities of AFM ordering in polar-stacked bilayer.

Figure 1 shows the schematics of NRSS in a polar-stacked AFM bilayer formed from two vdW ferromagnetic- (FM-) or AFM-ordered monolayers whose magnetic atoms lie in the same plane. We do not consider monolayers holding $[C_2||\tau]$ symmetry, since in this case, the polar stacking does not guarantee the occurrence of NRSS. To create a polar-stacked bilayer structure, first, a bilayer with $A\bar{A}$ stacking is formed by placing monolayer $\bar{A}$, which represents mirror reflection through the $xy$ plane of monolayer $A$, on top of the monolayer $A$ (Fig. 1(a)). Two magnetic sublattices of the $A\bar{A}$ bilayer are connected by symmetries $\{\bar{E}, M_z, C_{2z}\}$, which implies spin degeneracy of its electronic band structure. The $A\bar{A}$ state is however unstable and spontaneously relaxes to a stable $A\bar{B}$ or $\bar{B}A$ state through translation of the top monolayer along a high-symmetry direction by $\mathbf{t}_{||}$ or $-\mathbf{t}_{||}$, respectively (Figs. 3(b-g)). These new structures, $A\bar{B}$ and $\bar{B}A$, break inversion symmetry and, depending upon the lateral translation, $t_{||}$ or $-t_{||}$, produce out-of-plane electric polarization, $P$ or $-P$, respectively. The magnetic sublattices of $A\bar{B}$- and $\bar{B}A$-stacked bilayers are not connected by any symmetry $\{\bar{E}, M_z, C_{2z}\}$, lifting Kramers' spin degeneracy. Yet, the magnetic sublattices may or may not be connected by crystal rotation symmetry, resulting in altermagnetic or non-altermagnetic NRSS, respectively.

Depending on the AFM order of the polar-stacked bilayer in the $A\bar{B}$ and $\bar{B}A$ states, we categorize polar stacking into three types. In type-I and type-II polar stackings, the individual layers of a bilayer host an AFM order. In the type-I stacking, the AFM order of the top layer is the same as that of the bottom layer (Figs. 1(b, c)). In contrast, in the type-II stacking, the AFM order of the top layer is magnetically reversed (Figs. 1(d, e)). Type-III polar stacking consists of FM-ordered monolayers with AFM interlayer coupling, commonly known as an A-AFM order (Figs. 1(f, g)). Due to the intralayer AFM ordering, bilayers of type-I and type-II stackings do not reveal net magnetization, while the type-III stacking produces a small net magnetic moment which is dependent on the electric polarization direction, indicating a magnetoelectric effect [60, 61].

The opposite polarized states $A\bar{B}$ and $\bar{B}A$ are related by the crystal symmetry $M_z$. Since the spin and lattice are decoupled, NRSS is independent of $M_z$ and depends only on the AFM order of states $A\bar{B}$ and $\bar{B}A$. For the type-I stacking, the top and bottom layers have the same magnetic order, so states $A\bar{B}$ and $\bar{B}A$ are related by $[E||M_z]$ symmetry which does not change NRSS. For the type-II and type-III stackings, due to the reversed AFM order of the two layers, transformation of $A\bar{B}$ to $\bar{B}A$ must include spin-flip operation. As a result, the states with opposite electric polarization are related by $[C_2||M_z]$ symmetry which transforms $\varepsilon(s, k_x, k_y)$ to $\varepsilon(-s, k_x, k_y)$, and hence reverses NRSS and flips the uncompensated net magnetization.

Next, we consider specific examples of the different types of polar stacking using representative 2D vdW materials.

### B. Type-I polar-stacked AFM bilayer

For this type of stacking, we focus on transition-metal phosphorus trichalcogenides, $MPX_3$ ($M$ = Mn, Fe, Co, Ni; $X$ = S, Se), which have recently gained significant interest. Many of these compounds were successfully exfoliated into monolayers and revealed different types of AFM ordering [62-64]. In a $MPX_3$ monolayer (space group No. 162, $P\bar{3}1m$), six transition-metal atoms are arranged in a honeycomb structure surrounding the central $[P_2S_6]^{4-}$ cluster (Fig. 2(a)). Monolayer $MPX_3$ exhibits the Néel AFM order and thus holds $[C_2||\bar{E}]$ symmetry ensuring Kramers' degeneracy. A pristine bilayer $MPX_3$ (space group No. 147, $P\bar{3}$) is formed by the $AB$ layer stacking and also holds $[C_2||\bar{E}]$ symmetry. To produce NRSS, the $[C_2||\bar{E}]$ symmetry must be broken, while maintaining net magnetization zero. This is achieved by the polar stacking of a bilayer, as we have discussed above (Figs. 1(a-c)). To find the ground state of the polar-stacked MnPSe₃ bilayer, we calculate the total energy as a function of lateral shift of the top layer $\bar{A}$ (a mirror image of the bottom layer) with respect to the bottom layer $A$ along high symmetry directions [100] and $[1\bar{1}0]$. The calculations are performed as described in Methods (Sec. II). State $A\bar{A}$ has $[C_2||\bar{E}]$ symmetry, but as follows from our calculation (Fig. 2(e)), this structure is unstable. The lowest energy states $A\bar{B}$ and $\bar{B}A$ exhibit lateral shifts of layer $\bar{A}$ along the [100] direction by $a/3$ and $-a/3$, respectively, where $a$ is the lattice constant (Figs. 2(b, c)). The ground states $A\bar{B}$ and $\bar{B}A$ (space group No. 8, Cm), break mirror $M_z$ symmetry and produce electric polarization of 0.031 pC/m along $+z$ and $-z$ direction, respectively [55]. As seen from Figure 2(f), the ground-state energy is 0.1 meV lower for the type-I stacking than for the type-II stacking, and this difference remains unchanged upon including SOC. Thus, a polar-stacked MnPSe$_3$ bilayer belongs to the type-I category.

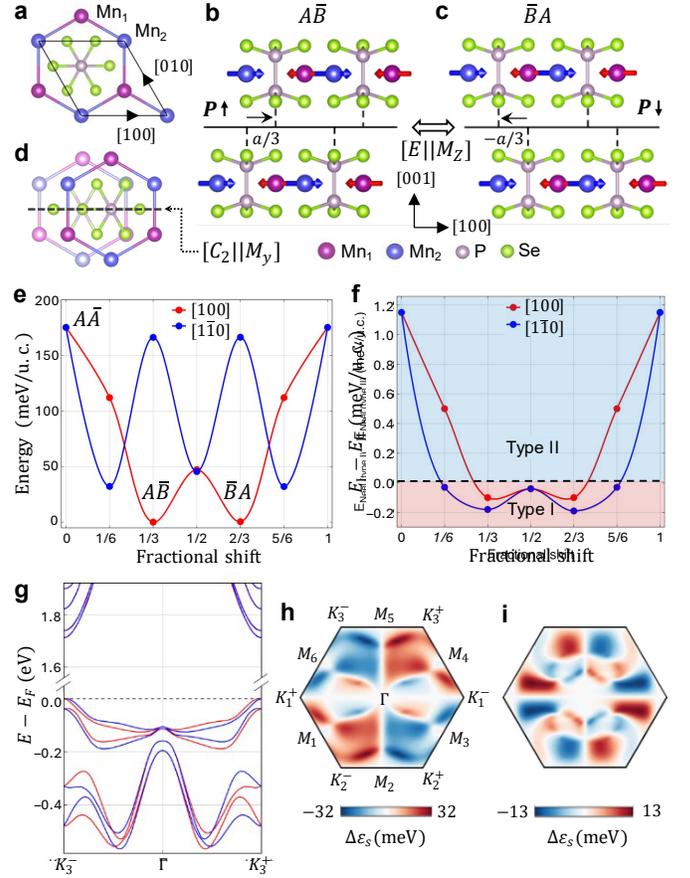

**FIG. 2: Results for type-I polar-stacked MnPSe₃ bilayer.** (a) Top view of a MnPSe₃ monolayer. Manganese atoms Mn₁ and Mn₂ represent two magnetic sublattices surrounding $[P_2S_6]^{4-}$ cluster. The hexagonal unit cell is indicated by black line. (b,c) Side view of the MnPSe₃ bilayer in ground states $A\bar{B}$ (c) and $\bar{B}A$ (d) related by $[E||M_z]$ symmetry. Electric polarization **P** is indicated by vertical arrows (d) Top view of the $A\bar{B}$ structure. Saturated (unsaturated) color denotes top (bottom) layer. $[C_2||M_y]$ symmetry is indicated by black line. (e) Relative stacking energy as function of lateral shift of the top layer along [100] and $[1\bar{1}0]$ directions. (f) Energy difference between type-I and type-II stackings as function of lateral shift of the top layer along [100] and $[1\bar{1}0]$ directions. (g) Spin-polarized band structure of the $A\bar{B}$-stacked bilayer. (h,i) NRSS energy $\Delta\varepsilon_s = \varepsilon(s, k_x, k_y) - \varepsilon(-s, k_x, k_y)$ of the highest valence band (i) and the lowest conduction band (j) of the $A\bar{B}$-stacked bilayer plotted in the full Brillouin zone.

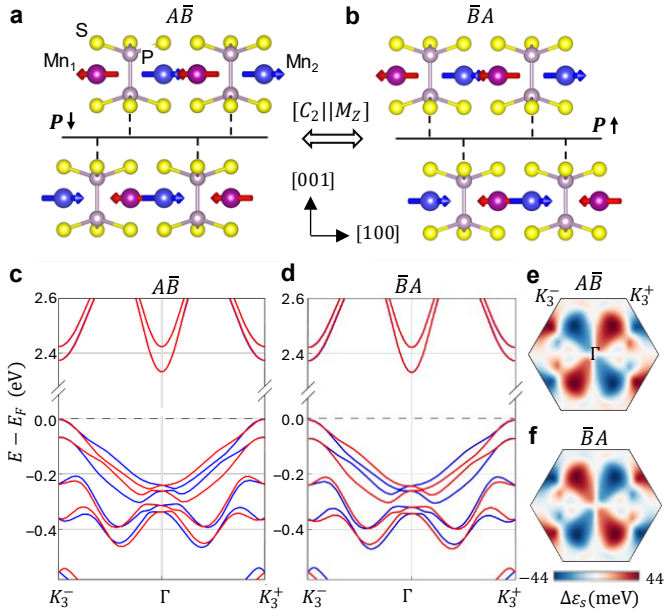

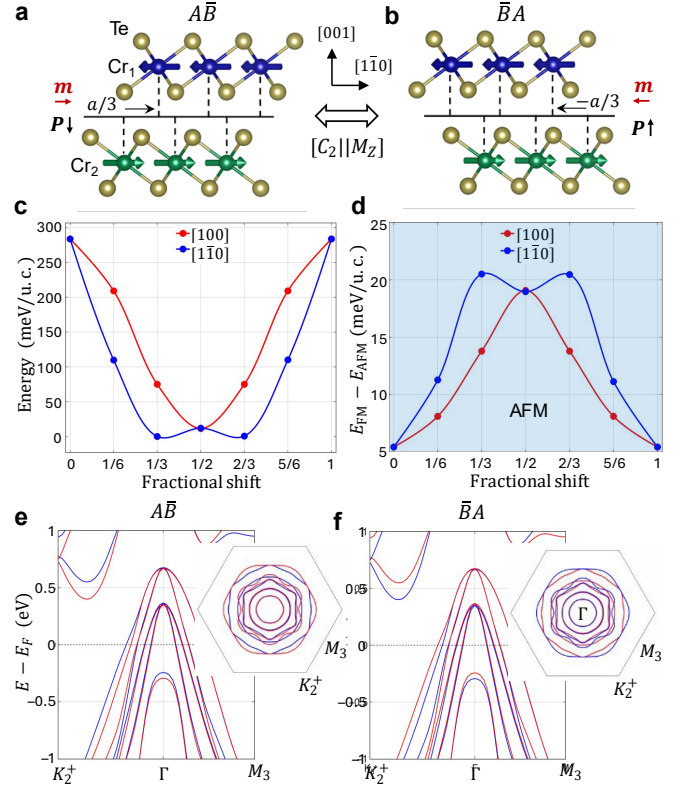

FIG. 3: Results for type-II polar-stacked 5% tensile strained MnPSe$_3$ bilayer. (a,b) Side view of the MnPSe$_3$ bilayer in ground states $A\bar{B}$ (c) and $\bar{B}A$ (d) related by $[E||M_z]$ symmetry. (c,d) Spin-polarized band structure of the $A\bar{B}$- (c) and $\bar{B}A$- (d) stacked bilayer along high-symmetry $\Gamma$ - $K_3^-$ and $\Gamma$ - $K_3^+$ paths. (e,f) NRSS energy $\Delta\varepsilon_s = \varepsilon(s,k_x,k_y) - \varepsilon(-s,k_x,k_y)$ of the highest valence band of the $A\bar{B}$- (e) and $\bar{B}A$- (f) stacked bilayer plotted in the full Brillouin zone.

The spin-space group of the polar ground states $A\bar{B}$ and $\bar{B}A$ hold symmetries $[E||E]$ and $[C_2||M_y]$. Two magnetic sublattices of the bilayer are interchanged by mirror symmetry $M_y$, not any of the symmetries $\{\tau, \bar{E}, M_z, C_{2z}\}$, making states $A\bar{B}$ and $\bar{B}A$ altermagnetic. As a result, at wave vectors $\mathbf{k}$ noninvariant to $M_y$, energy bands $\varepsilon(s,\mathbf{k})$ exhibit NRSS, e.g., along $\Gamma$-$K_3^-$ and $\Gamma$-$K_3^+$ paths shown in Fig. 2(g). $[C_2||M_y]$ and its combination with $[\bar{C}_2||T]$ transform $\varepsilon(s,\mathbf{k})$ as follows: $[C_2||M_y]\, \varepsilon(s,k_x,k_y) = \varepsilon(-s,k_x,-k_y)$, $[C_2||M_y][\bar{C}_2||T]\, \varepsilon(s,k_x,k_y) = \varepsilon(-s,-k_x,k_y)$. These symmetries enforce spin degeneracy at the $k_x = 0$ and $k_y = 0$ lines, as seen from NRSS energy $\Delta\varepsilon_s = \varepsilon(s,k_x,k_y) - \varepsilon(-s,k_x,k_y)$ of the highest valence band (Fig. 2(h)) and the lowest conduction band (Fig. 2(i)) in the full Brillouin zone. The two spin-degenerate nodal lines imply that the polar-stacked MnPSe$_3$ bilayer is a $d$-wave altermagnet [6]. We find maximum NRSSs of about 32 and 13 meV for the highest valence and lowest conduction bands, respectively. These values are larger than those previously reported for MnPSe$_3$ when it is intercalated [39] or twisted [65]. As a MnPSe$_3$ bilayer hosts the type-I ground state, reversal of its electric polarization does not change the NRSS sign, i.e., states $A\bar{B}$ and $\bar{B}A$ have the same NRSS.

### C. Type-II polar-stacked AFM bilayer

The spin-space group of the type-II polar stacking involves the same symmetry operations, $[E||E]$ and $[C_2||M_y]$, as the type-I stacking. Thus, any type-II polar-stacked $M$P$X_3$ bilayer with intralayer Néel AFM order exhibits $d$-wave altermagnetism. To demonstrate type-II polar stacking, we consider a strained MnPS$_3$ bilayer. Similar to MnPSe$_3$, MnPS$_3$ exhibits an intralayer Néel AFM order. We find that the type-I and type-II polar stackings of a MnPS$_3$ bilayer have almost the same energy. However, applying biaxial tensile strain of 4% and above forces a type-II polar ground state (see Fig. S1 in Supplemental Material [55]).

Figures 3(c, d) show the calculated band structures of the $A\bar{B}$- and $\bar{B}A$-stacked MnPS$_3$ bilayers exposed to 5% biaxial tensile strain. We find that the dipole moment of the strained MnPS$_3$ bilayer is about 0.27 pC/m, which is much larger than that of MnPSe$_3$ [55]. Contrary to the type-I stacking, the type-II stacking of the strained MnPS$_3$ bilayer forms states $A\bar{B}$ and $\bar{B}A$, which are related by $[C_2||M_z]$ symmetry transforming $\varepsilon(s, k_x, k_y)$ to $\varepsilon(-s, k_x, k_y)$ (Figs. 3(a, b)). As a result, these states have an opposite NRSS (Figs. 3(e,f)). Thus, reversing electric polarization of the MnPS$_3$ bilayer flips the NRSS sign, while preserving the Néel vector.

### D. Type-III polar-stacked AFM bilayer

Finally, we consider polar stacking of an A-AFM bilayer, where two FM-ordered monolayers are AFM coupled. Such A-AFM ordering has been observed in several vdW materials, such as CrTe$_2$, CrSBr, MnBi$_2$Te$_4$, etc. [66- 75]. The spin degeneracy in the pristine A-AFM bilayer is protected by $[C_2||\bar{E}]$ and/or $[C_2||M_z]$ symmetry. A polar-stacked A-AFM bilayer breaks these symmetries, producing NRSS. Since there are no symmetries connecting the two magnetic sublattices, this NRSS is non-altermagnetic [8]. In this case, the magnetization of the two FM-ordered layers is slightly different, creating a small net magnetic moment and spin splitting at the Γ point. Due to the two opposite polarization states in an A-AFM polar-stacked bilayer being related by $[C_2||M_z]$ symmetry (Figs. 1(f, g)), reversal of electric polarization changes the NRSS sign and flips the uncompensated net magnetic moment.

We demonstrate these properties by considering a polar-stacked 1T-CrTe$_2$ bilayer. In a pristine AB-stacked 1T-CrTe$_2$ bilayer, Kramers' spin degeneracy is protected by $[C_2||\bar{E}]$ symmetry and in an $A\bar{A}$-stacked bilayer, by $[C_2||M_z]$ symmetry. The $A\bar{A}$ stacking is unstable and relaxes to polar-stacked states, $A\bar{B}$ or $\bar{B}A$ (Figs. 4 (a,b)), through $\pm a/3$ lateral shifts of layer $\bar{A}$ along the [1$\bar{1}$0] direction (Fig. 4(c)). We find that the dipole moment of the polar-stacked 1T-CrTe$_2$ bilayer is about 0.20 pC/m [55]. The calculated energy difference between FM and AFM orderings indicates that the polar-stacked 1T-CrTe$_2$ bilayer favors an AFM interlayer coupling (Fig. 4(d)). The ground states $A\bar{B}$ and $\bar{B}A$ belong to space group No. 156 (P3m1), which has no symmetry operation to swap the two magnetic sublattices. As a result, for state $A\bar{B}$, we find that Cr atoms in the top and bottom 1T-CrTe$_2$ layers have different magnetic moments, $3.19\mu_B$ and $-3.17\mu_B$, respectively. The net magnetic moment of the bilayer is about $0.03\mu_B$ per Cr atom. It lies in the plane of the bilayer, as dictated by out-of-plane magnetic anisotropy of 6.24 meV, and is pointing in opposite directions for states $A\bar{B}$ and $\bar{B}A$. The spin-space group of the polar ground states includes symmetry $[E||C_{3z}]$, which, when combined with $[\bar{C}_2||T]$, results in $[E||C_{6z}]$, indicating sixfold rotational symmetry of the spin-polarized band structure. This symmetry analysis is consistent with the results of our DFT calculations shown in Figures 4(e, f), which reveal a reversable NRSS with a maximum value of 130 meV at the Fermi energy.

### IV. CONCLUSIONS

Overall, our results demonstrate that polar-stacked AFM bilayers exhibit NRSS which can have either altermagnetic (types I and II) or non-altermagnetic (type III) origins. Both support a reversable NRSS (types II and III) driven by electric polarization switching of the bilayer. These results are robust with respect to SOC (Sec. IV in Supplemental Material [55]). Switching of the NRSS by voltage, rather than by electric current, is beneficial for AFM spintronics. It allows the control of NRSS without reversal of the AFM Néel vector providing non-volatile performance assured by the stability of the two electrically polarized states. The proposed approach is expected to be more energy-efficient than that based on switching the Néel vector by spin-transfer or spin-orbit torques [30]. We emphasize that these conclusions are not limited to the vdW compounds considered in this work but are applicable to other AFM-ordered vdW materials exhibiting type-II or type-III stackings. Recent experiments on ferroelectric switching of polar-stacked vdW 3R-MoS$_2$ and BN bilayers have demonstrated ultrafast fatigue-free ferroelectric switching over $10^{11}$ cycles [76- 78]. Thus, electric polarization switching of the proposed polar-stacked NRSS AFM bilayers is feasible experimentally providing a promising platform for low-power spintronics.

### ACKNOWLEDGMENTS

This work was primarily supported by the grant number DE-SC0023140 funded by the U.S. Department of Energy, Office of Science, Basic Energy Sciences (H. M., K. H., E. Y. T). Partial support from the National Science Foundation through the EPSCoR RII Track-1 program (NSF Award OIA-2044049) and from the UNL Grand Challenges catalyst award "Quantum Approaches Addressing Global Threats" (K. H., K. S., E.Y.T.) is acknowledged. Computations were performed at the University of Nebraska Holland Computing Center.